\documentclass[aps,prf,reprint,amsmath,amssymb]{revtex4-2}

\usepackage{graphicx}
\usepackage{dcolumn}
\usepackage{bm}

\usepackage{import}
\usepackage{setspace}

\usepackage[utf8]{inputenc} 
\usepackage[T1]{fontenc}    
\usepackage{url}            
\usepackage{booktabs}       
\usepackage{amsfonts}       
\usepackage{nicefrac}       
\usepackage{microtype}      
\usepackage{amsthm}
\usepackage{subfigure}

\usepackage{capt-of}
\usepackage{epsfig} 
\usepackage{times} 
\usepackage{amsmath} 
\usepackage{amssymb}  
\usepackage{makeidx}
\usepackage{float}
\usepackage{balance}
\usepackage{booktabs}
\usepackage{bbm}
\usepackage{mathrsfs}
\usepackage{enumitem}
\usepackage{algorithm}
\usepackage{algpseudocode}

\usepackage[super]{nth} 

\usepackage[english]{babel} 

\usepackage{mathrsfs}  
\usepackage[normalem]{ulem}
\usepackage{xr} 
\usepackage{xcolor}
\usepackage{float}

\usepackage{xcolor}
\usepackage[colorlinks]{hyperref}       

\hypersetup{
  colorlinks = true,
  allcolors = siaminlinkcolor,
  urlcolor = siamexlinkcolor,
}
\colorlet{siaminlinkcolor}{blue!50!black}
\colorlet{siamexlinkcolor}{blue!50!black}

\usepackage{lipsum} 

\usepackage{setspace}

\usepackage{cleveref} 

\usepackage[acronym]{glossaries}

\usepackage{etoolbox}

\newcommand{\rd}{{\mathrm d}}

\newcommand{\vu}{{\bf u}}

\newcommand{\vH}{{\bf H}}

\newcommand{\vPhi}{{\mbox{\boldmath$\Phi$}}}

\newcommand{\vtheta}{{\mbox{\boldmath$\theta$}}}

\newcommand{\calD}{{\cal D}}

\newcommand{\calH}{{\cal H}}

\newcommand{\calL}{{\cal L}}

\newcommand{\calN}{{\cal N}}

\newcommand{\calU}{{\cal U}}

\newcommand{\argmax}{\operatornamewithlimits{argmax}}
\newcommand{\argmin}{\operatornamewithlimits{argmin}}

\newcommand{\Rb}{\mathbb{R}}

\newcommand{\tr}{{\mathrm {Tr}}}

\newcommand{\half}{\frac{1}{2}}

\newcommand{\Langle}{{\Big\langle}}
\newcommand{\Rangle}{{\Big\rangle}}

\makeatletter
\renewcommand{\maketag@@@}[1]{\hbox{\m@th\normalsize\normalfont#1}}%
\makeatother

\makeatletter
\newcommand{\vast}{\bBigg@{3.5}}
\newcommand{\Vast}{\bBigg@{4}}
\newcommand{\vastt}{\bBigg@{4.5}}
\newcommand{\Vastt}{\bBigg@{5}}
\newcommand{\vvast}{\bBigg@{5.5}}
\newcommand{\Vvast}{\bBigg@{6}}
\newcommand{\vvastt}{\bBigg@{6.5}}
\newcommand{\Vvastt}{\bBigg@{7}}
\makeatother

\setlist[enumerate]{
  label={\upshape(\roman*)},
  leftmargin=30pt,
  labelwidth=0pt,
  align=right,
  nosep
}

\makeglossaries
\newacronym{SPDE}{SPDE}{Stochastic Partial Differential Equation}
\newacronym{ODE}{ODE}{Ordinary Differential Equation}
\newacronym{PDE}{PDE}{Partial Differential Equation}
\newacronym{SDE}{SDE}{Stochastic Differential Equation}
\newacronym{POD}{POD}{Proper Orthogonal Decomposition}
\newacronym{MOR}{MOR}{Model Order Reduction}
\newacronym{MPC}{MPC}{Model Predictive Control}
\newacronym{RMSE}{RMSE}{Root Mean Squared Error}
\newacronym{HJB}{HJB}{Hamilton-Jacobi-Bellman}
\newacronym{ANN}{ANN}{Artificial Neural Network}
\newacronym{DNN}{DNN}{Deep Neural Network}
\newacronym{FNN}{FNN}{Feed-forward Neural Network}
\newacronym{CNN}{CNN}{Convolutional Neural Network}
\newacronym{LSTM}{LSTM}{Long-Short Term Memory}
\newacronym{FC}{FC}{Fully Connected}
\newacronym{PI}{PI}{Path Integral}
\newacronym{NN}{NN}{Neural Network}
\newacronym{SOC}{SOC}{Stochastic Optimal Control}
\newacronym{RL}{RL}{Reinforcement Learning}
\newacronym{RNN}{RNN}{Recurrent Neural Network}
\newacronym{DL}{DL}{Deep Learning}
\newacronym{SGD}{SGD}{Stochastic Gradient Descent}
\newacronym{IDVRL}{IDVRL}{Infinite Dimensional Variational Reinforcement Learning}
\newacronym{ROM}{ROM}{Reduced Order Model}
\newacronym{KL}{KL}{Kullback-Leibler}
\newacronym{DOF}{DOF}{Degrees of Freedom}
\newacronym{GRAPE}{GRAPE}{Gradient Ascent Pulse Engineering}
\newacronym{QND}{QND}{Quantum Non-Demolition}
\newacronym{LQR}{LQR}{Linear Quadratic Regulator}
\newacronym{1D}{1D}{1-dimensional}
\newacronym{2D}{2D}{2-dimensional}
\newacronym{3D}{3D}{3-dimensional}
\newacronym{RN}{RN}{Radon-Nikodym}
\newacronym{SNN}{SNN}{Sparse Neural Network}
\newacronym{FLOP}{FLOP}{Floating Point Operation}
\newacronym{GASS}{GASS}{Gradient-based Adaptive Stochastic Search}
\newacronym{NODE}{NODE}{Neural Ordinary Differential Equation}
\newacronym{QGASS}{QGASS}{Quantum Gradient-based Adaptive Stochastic Search}
\newacronym{MPPI}{MPPI}{Model Predictive Path Integral}
\newacronym{SME}{SME}{Stochastic Master Equation}
\newacronym{QRL}{QRL}{Quantum Reinforcement Learning}
\newacronym{ITC}{ITC}{Information Theoretic Control}
\newacronym{NP}{NP}{Non-Polynomial-Time}
\newacronym{ReLU}{ReLU}{Rectified Linear Unit}
\newacronym{MFC}{MFC}{Measurement-based Feedback Control}
\newacronym{CFC}{CFC}{Coherent Feedback Control}

\begin{document}

\preprint{APS/123-QED}

\title{Stochastic optimization for learning quantum state feedback control}

\author{Ethan N. Evans$^1$}%
\thanks{Now at the Naval Surface Warfare Center, Panama City Division, Panama City, FL 32407.}%
\thanks{Corresponding author. email: eevans89@gmail.com}%

\author{Ziyi Wang$^2$}%

\author{Adam G. Frim$^3$}%

\author{Michael R. DeWeese$^{3,4}$}%

\author{Evangelos A. Theodorou$^{1,2}$}%

\affiliation{%
  $^1$ Department of Aerospace Engineering, Georgia Institute of Technology \\
  Atlanta, GA, USA 30332
}%
\affiliation{%
  $^2$ Center for Machine Learning, Georgia Institute of Technology \\
  Atlanta, GA, USA 30332
}%
\affiliation{%
  $^3$ Department of Physics, University of California, Berkeley  \\
  Berkeley, CA, USA 94720
}%
\affiliation{%
  $^4$ Redwood Center for Theoretical Neuroscience and Helen Wills Neuroscience Institute, University of California, Berkeley  \\
  Berkeley, CA, USA 94720
}%


\begin{abstract}
\noindent \textbf{High fidelity state preparation represents a fundamental challenge in the application of quantum technology. While the majority of optimal control approaches use feedback to improve the controller, the controller itself often does not incorporate explicit state dependence.
Here, we present a general framework for training deep feedback networks for open quantum systems with quantum nondemolition measurement that allows a variety of system and control structures that are prohibitive by many other techniques and can in effect react to unmodeled effects through nonlinear filtering. We demonstrate that this method is efficient due to inherent parallelizability, robust to open system interactions, and outperforms landmark state feedback control results in simulation.}
\end{abstract}


\maketitle

\section{Introduction}

The efficacy of quantum technologies is fundamentally linked to our ability to prepare, stabilize, and steer between quantum states. Examples include gate synthesis and state preparation in quantum computing~\cite{schulte2005optimal,sporl2007optimal}, quantum metrology~\cite{chan2011laser}, quantum chemistry~\cite{maday2003new}, nuclear magnetic resonance~\cite{khaneja2003optimal,khaneja2005optimal}, and molecular physics~\cite{shapiro2012quantum}. Complex scenarios require rich tools from optimization and control theory which provide successful protocols with guarantees. The lens of optimal control theory and stochastic optimization provides numerous methodologies which cast such efforts as optimization problems. 

Quantum control algorithms stemming from optimal control theory or optimization theory are dominated by approaches that do not incorporate state feedback explicitly in the control law~\cite{boscain2002optimal,boscain2004nonisotropic,boscain2006time,carlini2006time,salamon2012optimal,boscain2014minimal,romano2014geometric,albertini2016time,szakacs1994locking,sola1998optimal,bartana2001laser,koch2004stabilization,palao2008protecting,caneva2009optimal,eitan2011optimal,kumar2011optimal,palao2013steering,ndong2014time,khaneja2005optimal,jager2014optimal,dong2008incoherent,niu2019universal}. These approaches produce a type of control known as open-loop control whose action is independent of the current system state and may perform well in certain circumstances where there are no unmodeled interactions or other effects. However, state feedback, i.e. where the control law explicitly depends on state information, is a primary tool of guaranteeing the stability of equilibria in the classical regime, where equivalent open-loop methods are often outperformed by their closed-loop counterparts in terms of robustness to unmodeled effects. 

Quantum control approaches that incorporate feedback are broken into two subcategories: \gls{MFC} and \gls{CFC}~\cite{zhang2017quantum}. In \gls{MFC}, state measurements are obtained through a classical measurement system coupling, which perturbs the system through a measurement back-action, and is used for control through an ancillary system coupling~\cite{doherty2000quantum,jacobs2003project,sayrin2011real,lu2017enhancing,cardona2018exponential,mulero2020quantum,warszawski2020solving,ma2021quantum}. On the other hand, \gls{CFC} designs a coherent coupling that controls the system state without measurement back-action~\cite{kashiwamura2017dispersive,goerz2018efficient,heuck2020controlled,krastanov2021controlled}. While \gls{CFC} methods may have advantages in terms of extracting system entropy~\cite{jacobs2014coherent}, \gls{MFC} methods can amplify measurement signals and apply macroscopic fields for feedback~\cite{balouchi2016coherent}.
\\
\indent
Within the \gls{MFC} setting, measurements cause a back-action onto the quantum systems, which typically causes a severe discontinuous jump into a system eigenstate, and as a result, are often reserved for post-experiment feedback. However, weak continuous \gls{QND} measurement protocols reduce the discontiuous back-action effect to a continuous Wiener diffusion process appended to the system state evolution, as originally suggested by Belavkin~\cite{belavkin1989new,belavkin1999measurement,belavkin1994nondemolition}. \gls{QND} measurement provides a partially observable state measurement which can be used throughout a given experiment for state feedback control. \gls{QND} measurement schemes have gained significant traction~\cite{milburn1983quantum,brune1990quantum,takahashi1999quantum,lupacscu2007quantum,nakajima2019quantum} and enable \gls{MFC} architectures that can effectively perform quantum feedback control on a variety of tasks~\cite{wiseman1993quantum,sayrin2011real,abdelhafez2019gradient}. \gls{QND} \gls{MFC} may yet hold the key to reducing the necessary qubit overhead in modern quantum computing architectures~\cite{ahn2002continuous} and has the promise of improving robustness and performance of many other quantum technologies.\\ \indent
In this paper, we apply optimization principles to closed loop state feedback control in a \gls{QND} \gls{MFC} setting. We leverage stochastic optimization and optimal control theoretic techniques, as well as tools from machine learning, to develop a general framework for learning control policies 
that perform feedback control for quantum state preparation and stabilization tasks. The approach is applied in simulation to a two-qubit stabilization task, and compared to a seminal state feedback \gls{QND} \gls{MFC} approach.

\section{Problem Formulation}

\begin{figure*}[t!]
\centering
    {\hspace{-0.15cm}\includegraphics[width=1.0\textwidth]{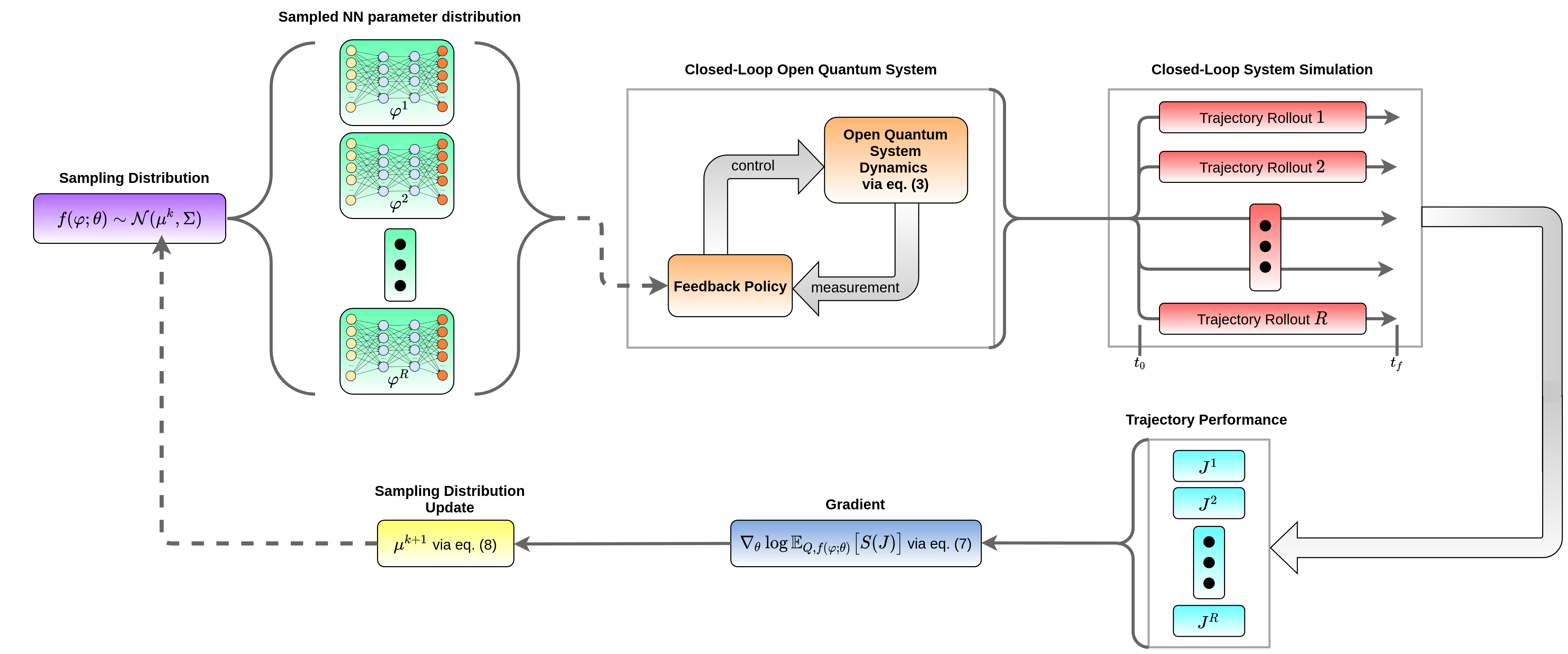}} 
\caption{Diagram of the QGASS policy learning architecture. A distribution of policy network parameters are sampled from sampling distribution $f(\varphi; \vtheta)$, which are used to generate a set of trajectory rollouts from the closed loop open quantum system dynamics described in \cref{eq:dyn_controlled}. The state output of the system is the result of a \gls{QND} measurement process as described in \cref{eq:continuous_belavkin}. The performance of each rollout is evaluated on $J(\rho, u)$ over the trajectory, which is then used to approximate the path integral expectation in the gradient \cref{eq:param_grad}. The gradient is then used to generate an update to the distribution for the next iteration, and iterations continue until convergence. Network parameter samples and trajectory rollouts are performed in parallel for computational efficiency.}
\label{fig:QGASS_diagram}
\end{figure*}

The \gls{QND} measurement scheme can be generally broken into discrete measurement (e.g. photon counting experiments), and continuous measurement (e.g. Homodyne and Heterodyne measurement experiements). The continuous measurement case yields dynamics driven by the Belavkin equation~\cite{belavkin1989new,belavkin1999measurement,belavkin1994nondemolition}, or  more generally a \gls{SME} of the form
\begin{align}
    \quad \rd \rho_t^c &= \calL_0 \rho_t^c \rd t + \calD[V]\rho_t^c \rd t \nonumber \\
    &\quad + \Big(V \rho_t^c+ \rho_t^c V^\dagger - \tr \big[(V + V^\dagger) \rho_t^c\big] \rho_t^c \Big) \rd W_t \label{eq:continuous_belavkin}
\end{align}
with innovation process
\begin{align}
    \rd W_t &= \rd y_t - \tr\big[(V + V^\dagger)\rho_t^c \big] \rd t \label{eq:continuous_belavkin_innovation}
\end{align}
where $\rd W_t$ is a standard zero-mean Wiener process in the classical sense~\cite{barchielli2009quantum} and $\rho_t^c$ is the system density state conditioned on the measurement outcome. This conditioning is key; if we were to "throw out" the measurement process, we would average over the noise process $\rd W_t$ and thus recover the deterministic Lindblad master equation~\cite{doherty2000quantum,barchielli2009quantum}. Here, the system closure includes the system $S$ and the measurement process $R$ with interaction operator $V$. One can similarly consider a closure that includes the system $S$, the environment $B$ and the measurement process $R$, however this is omitted for brevity.\\
\indent \Cref{eq:continuous_belavkin} is a quantum \gls{SPDE} with quantum unconditional evolution governed by the Lindblad terms $(\calL_0 + D[V])\rho_t^c \rd t$ and \gls{QND} measurement conditional evolution term $(V \rho_t^c+ \rho_t^c V^\dagger - \tr [(V + V^\dagger) \rho_t^c] \rho_t^c ) \rd W_t$. It is interesting to note that one can draw parallels between \cref{eq:continuous_belavkin} and the Kushner-Stratonovich \gls{SPDE}~\cite{doherty2000quantum}. Just as in the case of the Kushner-Stratonovich \gls{SPDE}, the stochasticity is the result of conditioning on the measurement process $\rd y_t$. Following this logic, one can think of the Belavkin equation in terms of a partially observable stochastic optimal control problem.\\
\indent
The open quantum system described by \cref{eq:continuous_belavkin} describes an \textit{uncontrolled} system. Control is typically introduced via a controlling potential described by a control Hamiltonian, which yields controlled open system dynamics given by
\begin{align}
    \rd \rho_t^c &= \calL_0 \rho_t^c \rd t + \calD[V] \rho_t^c - i \sum_j \vu_{t,j} [H_{u,j}, \rho_t^c] \rd t \nonumber \\ 
    &\quad + \Big(V \rho_t^c + \rho_t^c V^\dagger - \tr \big[(V + V^\dagger) \rho_t^c\big] \rho_t^c \Big) \rd W_t,
\end{align}
which can be equivalently expressed compactly by a simplified form
\begin{equation}\label{eq:belavkin_simp}
    \rd \rho_t^c = F(\rho_t^c) \rd t + G(\rho_t^c, \vu_t) \rd t + B(\rho_t^c) \rd W_t,
\end{equation}
where the term $F(\rho_t^c) \rd t := \calL_0 \rho_t^c \rd t + \calD[V] \rho_t^c$ describes the uncontrolled drift of the dynamics, the term $G(\rho_t^c, u_t) := - i \sum_j u_{t,j} [H_{u,j}, \rho_t^c] \rd t$ describes the controlled drift of the dynamics, and the term $B(\rho_t^c) \rd W_t := (V \rho_t^c + \rho_t^c V^\dagger - \tr [(V + V^\dagger) \rho_t^c] \rho_t^c ) \rd W_t$ describes the diffusion. Note that in the closed-loop control setting, the control $u$ has an explicit dependence on the state, i.e. $u = u( \rho_t^c)$, whereas in the open loop setting, it may be only time-dependent, i.e. $u=u(t)$.\\
\indent
A critical challenge in applying methods from stochastic optimal control is that in many cases, the operator-valued functional $B(\rho_t^c)$ can become singular, leading to a degenerate diffusion process. This is discussed with greater detail in the supplemental material, including some common instances of degeneracy. These degeneracies prove prohibitive for a variety of methods introduced in the stochastic optimal control literature, including path integral control~\cite{Kappen2005b,Kappen2016,williams2015model,williams2017model}, forward-backward stochastic differential equations using importance sampling~\cite{pereira2019learning,wang2019deep}, and recently spatio-temporal stochastic optimization~\cite{evans2020spatio,evans2021stochastic}. In each case, such degeneracies must be carefully addressed. In this paper, they are overcome by the proposed stochastic optimization technique.\\
\indent
The form of the dynamics in \cref{eq:belavkin_simp} is quite general and familiar in the context of optimal control theory. From this perspective, state preparation tasks are described in terms a positive-definite performance metric known as a cost functional $J(\rho_t^c, u_t)$, which typically uses a distance metric to penalize deviation from a target state and may additionally seek to reduce the control effort exerted onto the system. In many cases, the cost functional may be discontinuous or nondifferentiable (e.g. in the case of barrier functions or indicator functions), which can impose difficulties on control approaches. \\
\indent For concreteness, consider the task of reaching some target state $\rho_{\text{des}}$, as evaluated by the cost metric $J(\rho^c_t, u_t)$. The minimizing control is most generally expressed by the following path integral optimization problem
\begin{subequations}
\begin{align}
    \vu^* &= \argmin_{\vu \in \calU} \Langle J(\rho^c, u) \Rangle_Q \\
    \text{s.t. }\quad  \rd \rho_t^c &= F(\rho_t^c) \rd t + G(\rho_t^c, \vu_t) \rd t + B(\rho_t^c)\rd W_t, 
\end{align}
\end{subequations}
where the expectation defines a path integral over controlled state trajectories with path measure $Q$. The set $\calU$ is the admissible set of controls and may impose constraints on the control. One may also include constraints on the state $\rho^c_t$, however these are omitted from this derivation for simplicity.

\section{Quantum Gradient-based Adaptive Stochastic Search for Training Feedback Policies}

To solve this problem, we take an approach from stochastic optimization literature known as \gls{GASS}~\cite{zhou2014gradient}. The \gls{GASS} approach offers generality, as well as having guarantees of convergence and rate of convergence. This approach in essence manipulates the optimization problem by swapping the optimization variables from the control policy to the distribution parameters of the policy. This allows us to bypass discontinuities and non-differentiability in the dynamics and cost function. We provide details of this approach in the supplemental material. The resulting stochastic optimization problem takes the form 
\begin{subequations}\label{eq:nl_pol_ssopt}
\begin{align}
    \vtheta^* = \argmax_\vtheta \ln &\Langle S\Big( J(\rho^c, \vu)\Big) \Rangle_{Q,f(\varphi;\vtheta)}\\
    \text{s.t. }\quad   \rd \rho_t^c &= F(\rho_t) \rd t^c + G(\rho_t^c, \vu_t) \rd t + B(\rho_t^c)\rd W_t, \\
    \vu_{t} &= \vPhi(\rho_{t}^c; \varphi) \\
   \varphi &\sim f( \varphi ; \vtheta)
\end{align}
\end{subequations}
where the subscript on the expectation denotes a double expectation with respect to the path measure $Q$ of the controlled system dynamics, and with respect to some distribution $f$ belonging to the exponential family distributions. The function $S(\cdot)$ is a smooth, non-increasing shaping function and $\vPhi$ is a neural network which takes state information and outputs a control action. Such a neural network is typically referred to as a policy network, and in this case is dependent on a set $\varphi$ of weights and biases, which are sampled from distribution $f$ with parameters $\vtheta$. \gls{GASS} has been applied for a variety of optimization problems~\cite{zhou2017gradient, chen2018discrete, zhu2018simulation} and has also been explored in the context of optimal control~\cite{boutselis2020constrained, wang2021adaptive, exarchos2020novas}, however it is also appealing and pertinent in the context of policy learning as applied in this work. We denote this quantum feedback policy learning architecture \gls{QGASS}.
\begin{figure*}[t!] 
\hspace{-7.75cm}\textbf{a}  \hspace{8.75cm}\textbf{b}
\vspace{1em}

\centering
    {\hspace{-0.1cm}\includegraphics[width=1.02\columnwidth]{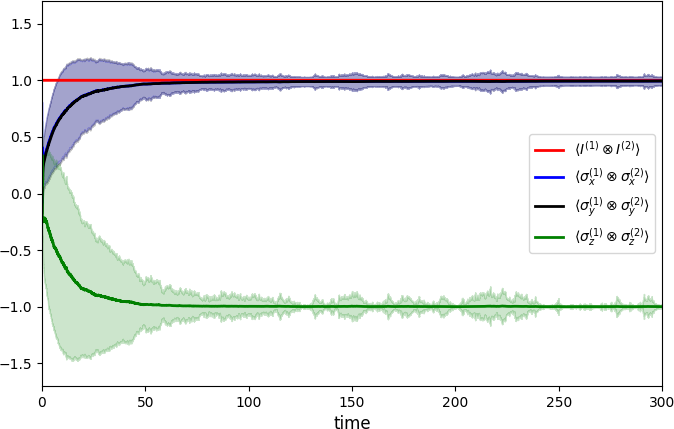}} {\hspace{0.2cm}\includegraphics[width=1.02\columnwidth]{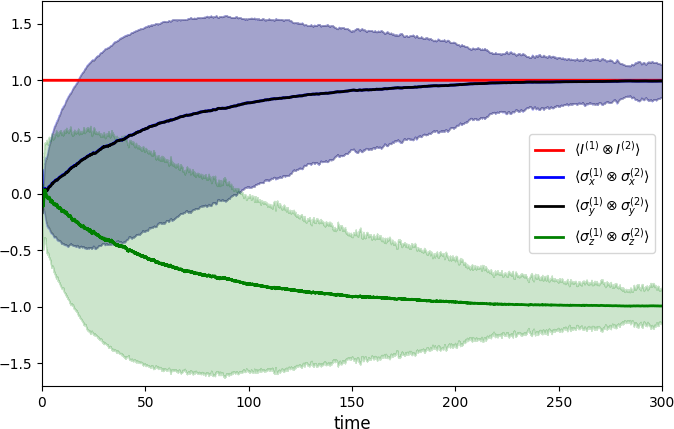}}

\caption{Two Qubit Symmetric Bell State Stabilization Task: \textbf{a}, Control with a linear policy trained by the QGASS algorithm. \textbf{b}, Control with the policy suggested by Mirrahimi, et. al. Colored lines denote the mean trajectories of the two qubit basis elements, and shaded regions denote 2-$\sigma$ variances. Means and variances are taken over 1000 trajectory rollouts.}
\label{fig:two_qubit_basis}
\end{figure*}
\\
\indent
As the name suggests, this approach performs a gradient-based update to the parameters of the sampling distribution $f(\varphi; \vtheta)$. Namely, the parameter gradient is obtained as 
\begin{align}
    \nabla_\vtheta \ln &\Langle S\big(J(\rho^c, \vu)\big) \Rangle_{Q, f(\varphi; \vtheta)} \nonumber \\
    &= \frac{\Langle S\big(J(\rho^c, \vu) \big) \big(T(x)-\nabla_\vtheta A(\vtheta)\big) \Rangle_{Q, f(\varphi; \vtheta)}}{\Langle S\big(J(\rho^c, \vu) \big)\Rangle_{Q, f(\varphi; \vtheta)}} \label{eq:param_grad}
\end{align}
where it is assumed that the sampling distribution $f(\varphi; \vtheta)$ belongs to the exponential family of distributions with sufficient statistics $T(x)$ and log partition function $A(\vtheta)$. Under a Gaussian sampling distribution $f(\varphi; \vtheta) \sim \calN(\mu, \Sigma)$, with mean update and fixed variance for simplicity, the parameter update becomes
\begin{align}
    \mu^{k+1} = \mu^{k} + \alpha^k\frac{\Langle S\big(J(\rho^c, \vu)\big) \big(\varphi-\mu^{k}) \Rangle_{Q, f(\varphi; \vtheta)}}{\Langle S\big(J(\rho^c, \vu)\big)\Rangle_{Q, f(\varphi; \vtheta)}}, \label{eq:pol_ss_param_update}
\end{align}

The \gls{QGASS} update scheme provides a parallelizable iterative training approach, which steers a distribution of network parameters towards learning optimal weights and biases, in turn providing a feedback control policy for the system. We include a detailed derivation of the \gls{QGASS} parameter update in the Supplemental material. Due to the path integral nature of this approach, the \gls{QGASS} approach is independent of discretization scheme used to discretize and simulate the dynamics in \cref{eq:dyn_controlled}. Furthermore, this approach can handle discontinuous jump-diffusion dynamics, such as in the discrete \gls{QND} measurement case~\cite{sayrin2011real}.

The \gls{QGASS} framework is shown in \cref{fig:QGASS_diagram}. To apply the algorithm, we first initialize the policy parameter distribution. Different policy realizations are then sampled from the distribution. For each policy sample, process noise is sampled to generate trajectory rollouts. The trajectory rollout propagation can be performed in parallel to significantly reduce runtime. The parameter update in \cref{eq:pol_ss_param_update} is performed through empirical approximation of the expectation using the rollout costs.

\section{Learning to Control Two-Qubits}

To illustrate the efficacy of the QGASS framework, we now demonstrate its use in practice. Specifically, we consider the control problem of stabilizing a two-qubit system to one of the Bell pair states of maximal entanglement. A stable solution to this problem was to shown to exist in the seminal work of Mirrahimi and Van Handel \cite{mirrahimi2007stabilizing}, though they did not consider the optimality of their solution. Here, we treat the symmetric up-down, down-up state as our target state. This is represented in the basis of two-qubit Pauli operators (i.e.  $\text{two qubit span} = \{I \otimes \sigma_x, I \otimes \sigma_y, I \otimes \sigma_z, \sigma_x \otimes I, \sigma_x \otimes \sigma_x, \sigma_x \otimes \sigma_y,\; \dots \}$). In this basis, the desired state can be represented as a linear combination of basis elements
\begin{align}
    \rho_{\text{desired}} &= \frac{1}{4} \big( I \otimes I + \sigma_x \otimes \sigma_x + \sigma_y \otimes \sigma_y - \sigma_z \otimes \sigma_z \big)
\end{align}
Thus, achieving the desired state can be viewed through expectations of these basis elements with respect to the conditioned density evolution\\
\indent
This experiment utilized a simulation environment built in Python, with state evolution adapted from the QuTip python library~\cite{johansson2012qutip}, and policy networks coded in PyTorch~\cite{paszke2019pytorch}. The algorithm computation speed is numerically improved by using vectorized (or batch) computations of the simulated trajectories, and CPU parellelization for policy parameter rollouts, resulting in $\sim$20 seconds per iteration for 1000 timesteps of an Euler-Maruyama discretization of \cref{eq:dyn_controlled} with 50 rollouts and 200 policy parameter rollouts. The algorithm was run on a desktop computer with an Intel Xeon 12-core CPU with a NVIDIA GeForce GTX 1060 GPU, and used less than 10 GB of RAM.\\
\indent
A \gls{FC} policy network was initialized via the method described in~\cite{lecun2012efficient}, and was trained for $850$ iterations of the \gls{QGASS} algorithm over a $1000$ timestep window. Despite being trained on just $1000$ timesteps of dynamics, the linear policy was tested on up to $100,000$ timesteps of dynamics and remained performant over the entire test window, in part due to the state feedback nature of the proposed control approach.\\
\indent
The trained policy network was then applied to an unseen test set of dynamics, and achieved quick stabilization convergence, as depicted in \cref{fig:two_qubit_basis}. The left subfigure depicts the \gls{QGASS} method, and the right subfigure depicts the method suggested by Mirrahimi, et. al.~\cite{mirrahimi2007stabilizing} under identical experiment parameters. Each solid line in each subfigure represents the mean expectation of the basis element, averaged over $1000$ test system trajectories, and the shading represents the 2-$\sigma$ variance of the distribution of expectations. In this basis, the \gls{QGASS} method can be observed to converge in approximately one order of magnitude faster than the benchmark, and has dramatically lower variance than the benchmark.\\
\begin{figure*}[t!] 
\hspace{-7.75cm}\textbf{a}  \hspace{8.5cm}\textbf{b}
\vspace{1em}
\centering
    {\hspace{-0.2cm}\includegraphics[width=2\columnwidth]{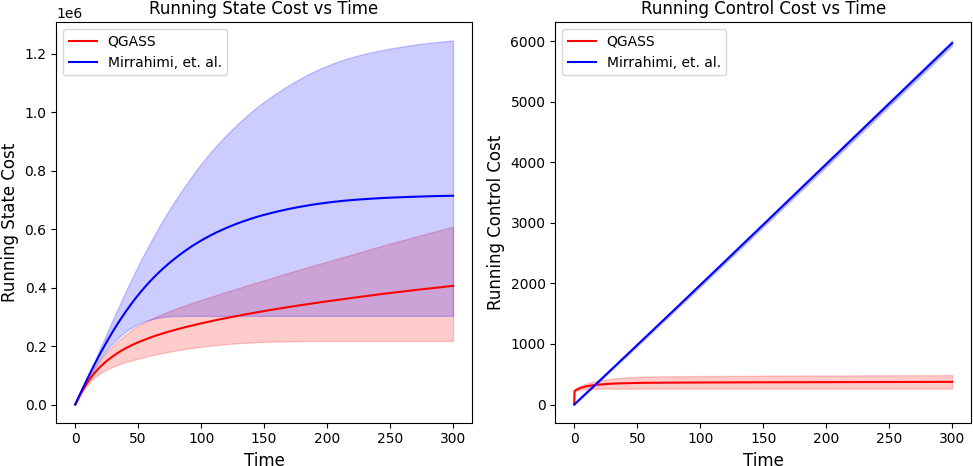}}

\caption{Two Qubit Symmetric Bell State Stabilization Task: \textbf{a}, running state costs and \textbf{b}, running control costs for the linear policy network trained with QGASS and the policy suggested by Mirrahimi, et. al. Colored lines denote the means and shaded regions denote 1-$\sigma$ variances. Means and variances are taken over 1000 sampled trajectories. The imposed data symmetry from this Gaussian depiction is corrected only when a large variance would lead to an infeasible negative instantaneous cost.}
\label{fig:two_qubit_costs}
\end{figure*}
\indent
The efficacy of the policy trained by \gls{QGASS} can also be visualized in terms of the cost metric $J(\rho_t^c, u_t)$. In \cref{fig:two_qubit_costs}, the running cost components of the policy trained by \gls{QGASS} are depicted. The left subfigure depicts the running state cost component of $J(\rho_t^c, u_t)$, given by
\begin{equation}
    J_{\text{state},t}(\rho_t) := \int_0^t Q_s q\big(1- \tr[\rho_{\text{desired}} \rho_\tau]\big) \rd \tau,
\end{equation}
while the right subfigure depicts the running control cost component of $J(\rho_t^c, u_t)$, given by
\begin{equation}
    J_{\text{control},t}(\vu) := \int_0^t \vu_\tau^\top Q_u \vu_\tau \rd \tau,
\end{equation}
where $Q_s$ and $Q_u$ are state cost and control cost weightings respectively, which are diagonal matrices weighing the cost along each state/control dimension. The solid line depicts the means of the running cost trajectories of each policy and the shading depicts the 1-$\sigma$ variance, each computed over $1000$ trajectory rollouts. The policy trained by \gls{QGASS} has a lower state cost on average, with a significantly lower 1-$\sigma$ variance of state cost, which suggests that the state performance, as measured by the running state cost metric, may have better guarantees of performance as compared to the policy suggested by Mirrahimi, et. al. The control effort of each policy is depicted in the right subfigure. It can be observed that the policy trained by \gls{QGASS} applies a strong initial control impulse to the system, followed by a relatively small control signal. This policy can be interpreted as a form of ``bang-bang" control. In contrast, the policy suggested by Mirrahimi, et. al. injects a fairly constant control signal over the time window, which yields a cumulative control effort which is approximately $12$ times higher than the policy trained by \gls{QGASS}.\\
\indent
Note that the impulsive control signal produced by the trained policy is likely to be experimentally realizable due to the viability of pulsed electromagnetic fields. However, if one were to desire a less impulsive control signal, one could add a running penalization term on the derivative of the control, effectively penalizing large rates of change in the control signal applied to the system~\cite{morzhin2019krotov}. One could also add a control rate indicator, effectively suppressing this additional cost until some control rate threshold is reached. This flexibility is possible since we do \textit{not} require any differentiability or even continuity of the cost functional in the \gls{QGASS} framework. While this will likely lead to a larger time-integrated control effort, one may either introduce terms to the cost functional or introduce hard constraints in order to enforce various experimental hardware constraints, including bounds on the control rate.

\section{Discussion \& Conclusion}

This paper suggests a method of training networks that utilize state information in an explicit state feedback control scheme. A key feature of such an explicit state dependence is the demonstrated extrapolation of the trained feedback scheme outside of the training window. State dependent feedback control can be contrasted against a multitude of existing methods that utilize a type of time-dependent feedforward control that is iteratively improved through measurement-based feedback. While omitting explicit feedback is often much simpler from a controller synthesis and optimization perspective, state feedback provides numerous benefits including stabilization performance over larger time scales.\\
\indent
Our results demonstrate efficacy of applying principles from stochastic optimization for \gls{QND} \gls{MFC}. We derived such a method, referred to as the \gls{QGASS} framework for training explicit feedback policy networks to control and stabilize the two-qubit experiment. The generality of the problem formulation suggests that this approach can be applied to the large class of experiments involving open quantum systems with either continuous \gls{QND} measurement schemes or discrete \gls{QND} measurement schemes. Furthermore, this approach is quite flexible; while we have only considered one form of cost functional and shape functional, there are virtually no limitations on the form of the cost functional, and there is quite a large variety of possible shape functionals. Combined with the fast computational speed of iterations, the results suggest that this approach can scale to larger numbers of qubits, which is actively being explored.\\
\indent
In~\cite{jacobs2013twenty}, the author suggests that \gls{CFC} subsumes \gls{MFC}, and suggests advantages of \gls{CFC} over \gls{MFC}~\cite{jacobs2014coherent}. In light of these remarks, one can generalize the \gls{QGASS} framework to train parametric coherent controllers coupled to the quantum system through implicit feedback (i.e. without an explicitly recorded measurement outcome). In such a case, the dynamics follow (deterministic) unconditional evolution governed by the Lindblad equation, forgoing a \gls{QND} measurement process entirely. Furthermore, we conjecture that through careful experiment design, it may be possible to apply \gls{QGASS} in a \gls{QND} \gls{MFC} setting to learn a parametric coherent controller, which is then applied to the system in a \gls{CFC} setting. \\
\indent
Our approach has similarities to the policy gradient method \cite{mulero2020quantum}, wherein gradients of the cost functional with respect to the control policy parameters are computed. However, the \gls{QGASS} approach is distinctive in that it assumes a probability distribution on the policy parameters and performs gradient updates on the distribution parameters instead. This can be seen as a form of abstraction, and allows one to bypass the often problematic differentiation steps through the cost functional and dynamics as in policy gradient methods. In addition, this means that our framework can be easily applied to problems with nondifferentiable or discontinuous dynamics and cost functions (e.g. photon counting and control thresholding).\\
\indent 
The \gls{QGASS} derivation for the policy parameter update holds true for arbitrary policy networks. In our simulated two-qubit control experiment, a rather shallow and simple linear feedback policy parameterization
is used.
An interesting next step is to investigate the effect of the size and depth of the policy network. 
The widespread success of utilizing deep networks for a variety of learning applications suggests that deeper network architectures may outperform the results presented here, especially for experiments with larger numbers of qubits. In addition, for more complex and higher dimensional experiments, convolutional or long-short term memory networks are also worth exploring to improve the scalability and temporal correlation of the policy.

\textbf{Acknowledgment} This work was supported by the Army Research Office contract W911NF2010151. ENE was supported by the Department of Defense though the SMART Scholarship program and the SMART SEED grant \href{www.smartscholarship.org}{www.smartscholarship.org}. ZW is supported by the NSFCPS award \#1932288. AGF is supported by the NSF GRFP under Grant No. DGE 1752814.





\bibliography{References}

\onecolumngrid\clearpage
\section*{Appendix A: Degenerate Diffusions in the Stochastic Master Equation}

The measurement-based feedback scheme for control of quantum systems has many advantages as outline in the main paper. However, this scheme also has certain pitfalls, which include problems of degeneracy of the diffusion dynamics. Degenerate diffusions can cause many control frameworks to fail, often due to the inability to find inverses or pseudoinverses of the covariance operator. In this section, we demonstrate two common examples in the context of quantum feedback control where this degeneracy may emerge due to a singular covariance operator.

\subsection{Degenerate Diffusions in the Two-Qubit system with QND Measurement}

Consider the two-qubit quantum system given by the stochastic master equation \cite{mirrahimi2007stabilizing}
\begin{equation}
    \rd \rho_t^c = -i u_{1}(t)[\sigma_y^{(1)}, \rho_t^c] \rd t - i u_2(t) [\sigma_y^{(2)}, \rho_t^c] \rd t - \half\big[F_z, [F_z, \rho_t^c]\big]\rd t + \sqrt{\eta}\Big( \{F_z, \rho_t^c\} - 2 \tr(F_z \rho_t^c) \rho_t^c \Big)\rd W_t,
\end{equation}
where $u_j(t)$ are two time-varying magnetic fields coupled to the two qubits, $F_z := \sigma_z^{(1)} \otimes I^{(2)} + I^{(1)} \otimes \sigma_z^{(2)}$ defines the coupling between the cavity and the electromagnetic field produced by the probe laser, as depicted in \cite[Figure 1.1]{mirrahimi2007stabilizing}, and as usual $[\cdot, \cdot]$ and $\{\cdot, \cdot\}$ are the commutator and anti-commutator, respectively. We can simplify this equation by defining the usual superoperators
\begin{align}
    \calH[A]\rho &:= \{A,\rho\} - 2 \tr(A \rho) \rho \label{eq:calH} \\
    \calD[A]\rho &:= \half\big[A, [A, \rho]\big] \label{eq:calD}
\end{align}
Also, note that in this system $H=0$, and $H_{u,j} = \sigma_y^{(j)}$. This yields
\begin{equation}
     \rd \rho_t^c = -\sum_{j} u_j(t)[H_{u_j}, \rho_t^c] \rd t - \calD[F_z] \rho_t^c \rd t + \sqrt{\eta}\calH[F_z] \rho_t^c \rd W_t
\end{equation}
so that we have the form in \cref{eq:belavkin_simp} repeated here for clarity
\begin{equation}
    \rd \rho_t^c = F(\rho_t^c) \rd t + G(\rho_t^c)\vu(t)\rd t + B( \rho_t^c) \rd W_t,
\end{equation}
where we have defined
\begin{align}
    F(\rho) &:= - \calD[F_z] \rho  \\
    G(\rho)u(t) &:= - i \sum_j u_j(t)\big[H_{u_j}, \rho \big] \\ 
    B(\rho) &:= \sqrt{\eta} \calH[F_z] \rho 
\end{align}

A key requirement of many stochastic optimal control methods is the invertability of the superoperator $\calH[\cdot]$, which can become singular. We can see this by simply looking at the $F_z$ operator. In this case it becomes 
\begin{align}
    F_z :\!\!&= I^{(1)} \otimes \sigma_z^{(2)} + \sigma_z^{(1)} \otimes I^{(2)} \nonumber \\
    &= \left[ \begin{array}{c c}
    \sigma_z & 0\\
    0 & \sigma_z
    \end{array} \right] + \left[ \begin{array}{c c}
    I & 0\\
    0 & -I
    \end{array} \right]  \nonumber \\
    &= \left[ \begin{array}{c c c c}
    1 & 0 & 0 & 0\\
    0 & -1 & 0 & 0 \\
    0 & 0 & 1 & 0 \\
    0 & 0 & 0 & -1
    \end{array} \right] + \left[ \begin{array}{c c c c}
    1 & 0 & 0 & 0\\
    0 & 1 & 0 & 0 \\
    0 & 0 & -1 & 0 \\
    0 & 0 & 0 & -1
    \end{array} \right]  \nonumber \\
     &= \left[ \begin{array}{c c c c}
    2 & 0 & 0 & 0\\
    0 & 0 & 0 & 0 \\
    0 & 0 & 0 & 0 \\
    0 & 0 & 0 & -2
    \end{array} \right]
\end{align}

Thus if the system is in the Bell state 
\begin{equation}
    \rho_{\text{desired}} = \left[ \begin{array}{cccc}
        0 &  0 & 0 & 0\\
        0 & 0.5 & 0.5 & 0 \\
        0 & 0.5 & 0.5 & 0 \\ 
        0 & 0 & 0 & 0
    \end{array}\right].
\end{equation},
the superoperator $\calH[F_z]\rho$ is a singular operator. The singularity above also arises if we rotate the magnetic fields such that they are coupled to the $x$ or $y$ axis of the spin representation, so that we have the coupling operator as $F_x := \sigma_x^{(1)} \otimes I^{(2)} + I^{(1)} \otimes \sigma_x^{(2)}$ or  $F_y := \sigma_y^{(1)} \otimes I^{(2)} + I^{(1)} \otimes \sigma_y^{(2)}$. In either of the three cases, the eigenvalues are $\lambda = \{0, 0, 2, 2 \}$. This is the case for \textit{any} $n$-qubit system.

\subsection{Degenerate Diffusions in the Homodyne QND Measurement Experiment}

The Homodyne detection experiment was among the first non-demolition measurement experiments, and can be viewed from the photon counting (jump noise) or continuous diffusion (Brownian noise) cases. In this experiment a cavity system emits photons when the atoms in the cavity are excited. The photon leakage is mixed with a local oscillator of the same frequency, and the mixed beam is then detected. The experimental setup is depicted in \cite[Figure A1]{verstraelen2018gaussian}.

The dynamics of the dissipative Homodyne detection experiment are given in Fock space by the stochastic master equation
\begin{equation}
    \rd \rho_t = -i [H_0, \rho_t] \rd t - i \sum_j u_j [H_{u_j}, \rho_t] \rd t - \half \sqrt{1-\eta} \sqrt{\gamma} \big[a, [a, \rho_t] \big] \rd t + \sqrt{\eta} \sqrt{\gamma} \Big( \{ a, \rho_t \} - 2 \tr(a \rho_t ) \rho_t \Big) \rd W_t
\end{equation}
where $H_0$ is the typical unforced Hamiltonian of the quantum harmonic oscillator, $a$ is the usual annihilation operator, and $\vH_u$ is the Hamiltionian of the external forcing, in this case provided by a coupled electromagnetic field. Using the previously defined $\calD$ and $\calH$ superoperators in \cref{eq:calD,eq:calH} yields the simplified form
\begin{equation}
     \rd \rho_t = -i [H_0, \rho_t] \rd t - i \sum_j u_j [H_{u_j}, \rho_t] \rd t -  \sqrt{1-\eta} \sqrt{\gamma} \calD[a]\rho_t \rd t + \sqrt{\eta} \sqrt{\gamma} \calH[a] \rho_t \rd W_t
\end{equation}
Again we have the form in \cref{eq:belavkin_simp}, with 
\begin{align}
    F(\rho_t) &:= -i[H_0, \rho_t] - \sqrt{1-\eta} \sqrt{\gamma}\calD[a]\rho_t \\
     G(\rho_t)u(t) &:= - i \sum_j u_j(t)\big[H_{u_j}, \rho_t \big] \\
    B(\rho_t) &:= \sqrt{\eta} \sqrt{\gamma} \calH[a] \rho_t
\end{align}

Investigating the $B(\rho_t)$ operator, we again find that it is singular, as can be seen by the form of the matrix representation of the annihilation operator $a$ for an $N$-level cavity
\begin{equation}
    a = \left[ \begin{array}{cccccc}
    0 & \sqrt{1} & 0 & \cdots && 0 \\
    0 & 0 & \sqrt{2} & 0 & \cdots & 0 \\
    \vdots && \ddots &&& \vdots \\
    0 && \cdots && 0 & \sqrt{N} \\
    0 && \cdots &&& 0
    \end{array} \right],
\end{equation}
which again leads to a singular superoperator for certain states. This singularity also arises if we use the creation operator $a^\dagger$ as the coupling operator.

\section*{Appendix B: QGASS Formulations for Learning Feedback Policies}

The \gls{GASS} method was introduced in~\cite{zhou2014gradient}, and has recently been applied as a control optimization strategy (c.f.~\cite{wang2021variational}).
This approach has provable convergence characteristics, and offers generality and flexibility. In this section, we will demonstrate this flexibility by exploring several problem formulations that leverage the approach for training policy networks for feedback control.

Consider the two qubit \gls{SME} in the general simplified form
\begin{equation} \label{eq:dyn_controlled}
    \rd \rho_t = F(\rho_t) \rd t + G(\rho_t, \vu_t) \rd t + B(\rho_t) \rd W_t
\end{equation}
where $G(\rho_t, \vu_t)$ is a state dependent controlled drift term. In the two-qubit problem $G(\rho_t, \vu_t)$ takes the form $G(\rho_t, \vu_t) = \sum_i^2 u_i [\sigma_y^{(i)}, \rho_t]$, however in a more general $N$-qubit experiment, one may require all single-particle Pauli matrices. Thus $G(\rho_t, \vu_t)$ may have the more general form
\begin{equation}
    G(\rho_t, \vu_t) = \sum_{i,j=1}^{N,3} u_{ij} \big[\sigma_{j}^{(i)}, \rho_t \big]
\end{equation}
where $\sigma_j^{(i)}$, $j \in {1,2,3}$ denote single particle Pauli matrices of each axis $x,y,z$. Despite appearing the context of qubit systems, the form of \cref{eq:dyn_controlled} is quite general, and can represent virtually \textit{any} open quantum system with continuous \gls{QND} measurement.

Quantum control problems often consider the task of reaching some target state $\rho_{\text{des}}$, as measured by some general cost metric $J(\rho_t, \vu_t)$. The minimizing control is most generally expressed by the following path integral optimization problem
\begin{subequations}
\begin{align}
    \vu^* &= \argmin_{\vu \in \calU} \Langle J(\rho, \vu) \Rangle_Q \\
    \text{s.t. }\quad  \rd \rho_t &= F(\rho_t) \rd t + G(\rho_t, \vu_t) \rd t + B(\rho_t)\rd W_t, 
\end{align}
\end{subequations}
where the expectation defines a path integral over controlled state trajectories with measure $Q$. The set $\calU$ is the admissible set of controls and may impose constraints on the control. One may also include constraints on the state $\rho$, however these are omitted from this derivation for simplicity.

The cost functional $J:H^2 \times \Rb^m \rightarrow \Rb$ is some real-valued, potentially non-convex, discontinuous, and non-differentiable functional, which must be minimized. Such a function imposes many difficulties from the context of optimization theory and optimal control theory. In the \gls{GASS} approach, we bypass these difficulties through stochastic approximation. Let $f(u;\vtheta)$ be a distribution belonging to the exponential family of distributions. Then the optimization problem is approximated as
\begin{subequations}
\begin{align}
    \theta^* &= \argmin_{\vtheta} \Langle J(\rho, \vu) \Rangle_{Q, f(\vu; \vtheta)} \\
    \text{s.t. }\quad  \rd \rho_t &= F(\rho_t) \rd t + G(\rho_t, \vu_t) \rd t + B(\rho_t)\rd W_t, \\
    \vu_{t} &\sim f(\vu_{t}; \vtheta)
\end{align}
\end{subequations}
Furthermore, we introduce the smooth (continuously differentiable a.e.), non-increasing shape function $S:\Rb \to \Rb$ and the logarithm function to obtain the following modified optimization problem 
\begin{subequations}\label{eq:open_loop_ssopt}
\begin{align}
    \theta^* &= \argmax_{\vtheta}\log \Langle S\big(J(\rho, \vu)\big) \Rangle_{Q,f(u;\vtheta)} \\
    \text{s.t. }\quad  \rd \rho_t &= F(\rho_t) \rd t + G(\rho_t, \vu_t) \rd t + B(\rho_t)\rd W_t, \\
    \vu_{t} &\sim f(\vu_{t}; \vtheta)
\end{align}
\end{subequations}
Solving this optimization problem with gradient-based parameter adaptation has been shown to have numerous appealing convergence characteristics detailed in~\cite{zhou2014gradient}, however a key observation is that this formulation does not incorporate the measurement from the measurement process $\rd W$ and is a purely feed-forward control. In this representation, one may compare this framework to the popular feedforward frameworks such as GRAPE or Krotov for optimal control of quantum systems without state feedback (c.f. the approaches in~\cite{jager2014optimal}), however, the goal in defining the \gls{SME} in \cref{eq:dyn_controlled} is to realize an explicit state feedback control optimization algorithm. In the following we consider a number of modifications to the above optimization problem to achieve this goal, each able to leverage the \gls{QGASS} training approach as outlined in the main paper.


\subsubsection{SME with linear parametric state feedback compensation.}
Consider the optimization problem
\begin{subequations} \label{eq:lin_pol_ssopt}
\begin{align}
    \vtheta^* = \argmax_\vtheta \log &\Langle S\Big( J(\rho, \vu)\Big) \Rangle_{Q,f(\varphi;\vtheta)} \label{eq:lin_pol_a}\\
    \text{s.t. }\quad   \rd \rho_t &= F(\rho_t) \rd t + G(\rho_t, \vu_t) \rd t + B(\rho_t)\rd W_t, \\
    \vu_t &= K_1(\varphi_1) \rho_t + K_2(\varphi_2) \\
   \varphi &:= [\varphi_1, \varphi_2] \sim f( \varphi ; \vtheta)
\end{align}
\end{subequations}
Under the realization that the controller in~\cite{mirrahimi2007stabilizing} is quite similar to a P-controller on the trace distance to the goal state, this has a static compensator with an explicit parametric linear feedback policy. The expectation in \cref{eq:lin_pol_a} is a double expectation composed of an expectation over the \gls{SME} and an expectation over the exponential family.

Note that this control policy can be realized through a fully connected network with \gls{ReLU} activations as
\begin{equation}
    \vu_t = K(\rho_t; \varphi),
\end{equation}
where $K: H^2 \to \Rb^m$ is the linear policy network. This motivates the use of nonlinear policy networks.

\subsubsection{SME with nonlinear parametric state feedback compensation.} Consider the optimization problem
\begin{subequations}\label{supeq:nl_pol_ssopt}
\begin{align}
    \vtheta^* = \argmax_\vtheta \log &\Langle S\Big( J(\rho, \vu)\Big) \Rangle_{Q,f(\varphi;\vtheta)} \\
    \text{s.t. }\quad   \rd \rho_t &= F(\rho_t) \rd t + G(\rho_t, \vu_t) \rd t + B(\rho_t)\rd W_t, \\
    \vu_{t} &= \vPhi(\rho_{t}; \varphi) \\
   \varphi &\sim f( \varphi ; \vtheta)
\end{align}
\end{subequations}
where $\vPhi$ is a nonlinear feedback policy parametrized by $\varphi$. This could be a \gls{FC} network or a \gls{CNN}, but in general simply represents a nonlinear function of $\rho$ without explicit time-dependence.

\subsubsection{SME with nonlinear recurrent state feedback compensation.} Consider the optimization problem
\begin{subequations}\label{eq:rnn_pol_ssopt}
\begin{align}
    \vtheta^* = \argmax_\vtheta \log &\Langle S\Big( J(\rho,\vu)\Big) \Rangle_{Q,f(\varphi,\vtheta)} \\
    \text{s.t. }\quad \rd \rho_t &= F(\rho_t) \rd t + G(\rho_t, \vu_t) \rd t + B(\rho_t)\rd W_t, \\
    \vu_{t_{k+1}} &= \vPhi_{RNN}(\rho_{t_k}, \vu_{t_k}; \varphi ) \label{eq:rnn_net_c}\\
    \varphi &\sim f( \varphi; \vtheta)
\end{align}
\end{subequations}
where $\vPhi_{RNN}$ is an RNN (e.g. LSTM network). Incorporating time-dependence in the policy endows the compensator with "dynamics", and enables treatment of a larger class of problems compared to a static compensator. 

One may also apply a \gls{NODE} network~\cite{chen2018neural} in place of \cref{eq:rnn_net_c}. Instead of specifying a discrete sequence of hidden layers, \gls{NODE} networks parametrize the derivative of the hidden state using a neural network, and as a result demonstrate \textit{constant} memory cost as a function of network depth, significantly lower training losses, and can handle time irregularity in the discretization scheme. In many cases, \gls{NODE} networks outperform \gls{RNN} networks, and are a closer representation to a dynamic compensation approach.

\subsubsection{SME with stochastic actuators and dynamic feedback compensation.} Consider the optimization problem
\begin{subequations} \label{eq:stoch_act_ssopt}
\begin{align}
    \vtheta^* = \argmax_\vtheta \log &\Langle S\Big( J(\rho, \vu)\Big) \Rangle_{Q,U,f(\varphi;\vtheta)} \\
    \text{s.t. }\quad  \rd \rho_t &= F(\rho_t) \rd t + G(\rho_t, \vu_t) \rd t + B(\rho_t)\rd W_t, \\
    \rd \vu_t &= G_u(\rho_t, \vu_t; \varphi_1) + \Sigma \rd V_{t}  \label{eq:stoch_act_ssopt_dyn_comp}\\
    \vu_{t_0} &= G_0(\varphi_2) \\
   \varphi &:= [\varphi_1, \varphi_2] \sim f( \varphi; \vtheta),
\end{align}
\end{subequations}
where $Q$ is the measure of the controlled dynamics, $U$ is the measure of the dynamic compensator, and $f( \varphi; \vtheta)$ is a distribution,  parameterized by $\vtheta$, which belongs to the exponential family of distributions. We include noise in the compensator to represent a realistic noisy digital compensation signal, however this can be neglected to reduce the sampling complexity. The function $J:H^2 \times \Rb^m \rightarrow \Rb$ is some real-valued, potentially non-convex and non-differentiable metric, and the function $S:\Rb \to \Rb$ is a smooth shape function. The function $G_u:H^2 \times \Rb^m \times \Rb^p$ is the drift of the dynamic compensator. 

Here, we must approximate the expectation with finite samples from three processes, namely the original \gls{SME}, the stochastic dynamic compensator, and the compensator initial condition distribution. This approach may enable substantially more exploration of the state space, however this comes at the cost of sampling \textit{three} distributions, which can quickly become computationally expensive. One may notice that these compensator dynamics are functionally similar to a stochastic \gls{RNN}.

\section*{Appendix C: QGASS Parameter Update Derivation}

The \gls{GASS} method was first derived in~\cite{zhou2014gradient}. Here we derive the parameter update under a minimization problem instead of a maximization problem, and use the above notation. Start with the general optimization problem
\begin{equation}
    \vu^* = \argmin_{\vu \in \calU} J(\rho, \vu)
\end{equation}
where $\calU \subseteq \Rb^n$ is a non-empty compact set, and $J: H \times \calU \to \Rb$ is a real-valued, potentially non-convex, discontinuous, and/or non-differentiable function. We avoid the inherent difficulties in $F(\vu)$ by transforming the problem into an approximation where $\vu$ is sampled from the distribution $f(\vu;\vtheta)$
\begin{equation}
    \vtheta^* = \argmin_{\vtheta} \int_\calU J(\rho, \vu) f(\vu;\vtheta) \rd \vu = \Langle J(\rho, \vu) \Rangle_{f(\vu; \vtheta)}
\end{equation}
The new problem formulation optimizes with respect to an upper bound of the original one since $\Langle J(\rho, \vu)\Rangle_{f(\vu;\vtheta^*)}\geq J(\rho, \vu^*)$. Equality is achieved when all the probability mass of $f(\vu;\vtheta^*)$ is at $\vu^*$. To facilitate the derivation, we additionally introduce a logarithmic transform and a shape function that is differentiable and non-increasing, $S(\cdot):\Rb \to \Rb_+$, which transforms our minimization problem into a maximization one
\begin{align}
    \vtheta^* &= \argmax_\vtheta \log \int_\calU S\big(J(\rho, \vu)\big) f(\vu;\vtheta) \rd \vu \\
    &= \log \Langle S\big(J(\rho, \vu)\big) \Rangle_{f(\vu; \vtheta)}
\end{align}

We perform gradient updates in order to update the parameters $\vtheta$ of the distribution $f(\vu;\vtheta)$, which is assumed to belong to the exponential family of distributions.
\begin{align}
    \nabla_\vtheta \log \int_\calU  S\big(J(\rho, \vu)\big) f(\vu;\vtheta) \rd \vu &= \frac{\nabla_\vtheta \int_\calU  S\big(J(\rho, \vu)\big) f(\vu;\vtheta) \rd \vu}{\int_\calU  S\big(J(\rho, \vu)\big) f(\vu;\vtheta) \rd \vu} \\
    &= \frac{ \int_\calU  S\big(J(\rho, \vu)\big) \nabla_\vtheta f(\vu;\vtheta) \rd \vu}{\int_\calU  S\big(J(\rho, \vu)\big) f(\vu;\vtheta) \rd \vu} \label{eq:gass_pass_grad}
\end{align}
Now we apply the log trick $\nabla_\vtheta f(\vu;\vtheta) = f(\vu;\vtheta) \nabla_\vtheta \log f(\vu;\vtheta)$ to obtain
\begin{align}
    \nabla_\vtheta \log \int_\calU  S\big(J(\rho, \vu)\big) f(\vu;\vtheta) \rd \vu &= \frac{ \int_\calU  S\big(J(\rho, \vu)\big) f(\vu;\vtheta) \nabla_\vtheta \log f(\vu;\vtheta) \rd \vu}{\int_\calU  S\big(J(\rho, \vu)\big) f(\vu;\vtheta) \rd \vu} \\
    &= \frac{\Langle S\big(J(\rho, \vu)\big) \nabla_\rho \log f(\vu;\vtheta) \Rangle_{f(\vu; \vtheta)}}{\Langle S\big(J(\rho, \vu)\big)\Rangle_{f(\vu; \vtheta)}}
\end{align}

The exponential family distribution is given by 
\begin{equation}
    f(\vu;\vtheta) = h(\vu)\big(\vtheta^\top T(\vu) - A(\vtheta) \big),
\end{equation}
which is characterized by a set of natural parameters $\vtheta$, sufficient statistics $T(\vu)$, base measure $h(\vu)$, and a log partition function $A(\vtheta)$. Thus we have
\begin{align}
    \nabla_\vtheta \log f(\vu;\vtheta) &= \nabla_\vtheta \log \Big[ h(\vu) \exp\big(\vtheta^\top T(\vu) - A(\vtheta) \big) \Big] \\
    &= \nabla_\vtheta \log h(\vu) + \nabla_\vtheta \big(\vtheta^\top T(\vu) - A(\vtheta) \big) \\
    &= T(\vu) - \nabla_\vtheta A(\vtheta)
\end{align}

If one optimizes only over the mean of a Gaussian distribution, then one obtains
\begin{align}
    \nabla_\vtheta \log \Langle S\big(J(\rho, \vu)\big) f(\vu;\vtheta) \Rangle_{f(\vu; \vtheta)} &= \frac{\Langle S\big(J(\rho, \vu)\big) \Sigma^{-1} \big(\vu-\mu) \Rangle_{f(\vu; \vtheta)}}{\Langle S\big(J(\rho, \vu)\big)\Rangle_{f(\vu; \vtheta)}},
\end{align}
where $\mu$ is the mean, and $\Sigma$ is the variance. Thus the gradient-ascent parameter update becomes
\begin{align}
    \Sigma^{-1}\mu^{k+1} = \Sigma^{-1}\mu^{k} + \Sigma^{-1} \frac{\Langle S\big(J(\rho, \vu)\big) \big(\vu-\mu) \Rangle_{f(\vu; \vtheta)} }{\Langle S\big(J(\rho, \vu)\big)\Rangle_{f(\vu; \vtheta)} }
\end{align}
or more simply
\begin{align}
    \mu^{k+1} = \mu^{k} + \frac{\Langle S\big(J(\rho, \vu)\big) \big(\vu-\mu^{k}) \Rangle_{f(\vu; \vtheta)} }{\Langle S\big(J(\rho, \vu)\big)\Rangle_{f(\vu; \vtheta)} }
\end{align}

Note that in the cases where we have added a level of abstraction due to the inclusion of a parameterized policy network $\Phi(\rho_t; \varphi)$, the above derivation yields a parameter update
\begin{align}
    \mu^{k+1} = \mu^{k} + \frac{\Langle S\big(J(\rho, \vu)\big) \big(\varphi-\mu^{k}) \Rangle_{f(\varphi; \vtheta)}}{\Langle S\big(J(\rho, \vu)\big)\Rangle_{f(\varphi; \vtheta)}}, \label{supeq:pol_ss_param_update}
\end{align}
where in this case $\mu$ is the mean of a Gaussian distribution on the policy network parameters $\varphi$.

Due to the path integral nature of this derivation, the so called \gls{QGASS} approach is independent of discretization scheme used to discretize the dynamics in \cref{eq:dyn_controlled}, as in the quantum trajectories literature. Furthermore, this approach may consider jump-diffusion dynamics, such as in the discrete measurement case~\cite{sayrin2011real}. 

Several of the above optimization problems contain two or three expectations, which is quite different than the above case wherein the parameter update was derived. In order to apply this parameter update to the two and three expectation cases above, one must simply re-define the shape function. In the cases of \cref{eq:open_loop_ssopt,eq:lin_pol_ssopt,supeq:nl_pol_ssopt,eq:rnn_pol_ssopt}, let the function $S(\cdot)$ be defined as
\begin{align}
    S(\cdot) := \Langle \hat{S}(\cdot) \Rangle_Q
\end{align}
where $\hat{S}$ is a standard shape function which is differentiable and non-increasing. Thus $S$ is non-increasing and positive semi-definite, so it may be treated as a shape function. This shape function may be substituted into \cref{supeq:pol_ss_param_update} to yield
\begin{equation}\label{eq:pol_ss_param_update_expectation}
    \mu^{k+1} = \mu^{k} + \frac{\bigg\langle \Langle \hat{S}\big(J(\rho, \vu)\big)\Rangle_Q \big(\varphi-\mu\big ) \bigg\rangle_{f(\varphi; \vtheta)}}{\bigg\langle \Langle \hat{S}\big(J(\rho, \vu)\big)\Rangle_Q \bigg\rangle_{f(\varphi; \vtheta)}}.
\end{equation}

Similarly, for \cref{eq:stoch_act_ssopt}, let the function $S(\cdot)$ be defined as
\begin{equation}
    S(\cdot) := \bigg\langle \Langle \hat{S}(\cdot) \Rangle_U \bigg\rangle_Q.
\end{equation}
In this case $S(\cdot)$ is also non-increasing and differentiable, so it may be treated as a shape function. This results in the parameter update
\begin{equation}\label{eq:stoch_act_ss_param_update}
     \mu^{k+1} = \mu^{k} + \frac{\Bigg\langle \bigg\langle \Langle \hat{S}\big(J(\rho, \vu)\big)\Rangle_U \bigg\rangle_Q \big(\varphi-\mu) \Bigg\rangle_{f(\varphi; \vtheta)}}{\Bigg\langle \bigg\langle \Langle \hat{S}\big(J(\rho, \vu)\big)\Rangle_U \bigg\rangle_Q \Bigg\rangle_{f(\varphi; \vtheta)}}.
\end{equation}

The parameter update in \cref{supeq:pol_ss_param_update} can be connected to the information theoretic version of the \gls{MPPI} algorithm for classical systems~\cite{williams2017information}, as first explored in~\cite{wang2021adaptive}. The information theoretic \gls{MPPI} algorithm applies an exponential shape function $S(y; \kappa) := \exp(-\kappa y)$ for $y,\kappa \in \Rb$, however other shape functions, such as the sigmoid function, are explored in~\cite{wang2021adaptive}.

The key difference between the \gls{QGASS} approach compared to \gls{MPPI}~\cite{williams2017information} is that \gls{MPPI} requires one to perform importance sampling, which presents challenges when the diffusion process becomes degenerate. Namely, the change of measures between the controlled and uncontrolled open quantum systems with \gls{QND} measurement requires inversion of an operator that is singular in a multitude of realizable experiments, such as the two-qubit system and the homodyne system.

In the context of~\cite{evans2021stochastic}, policies without explicit time dependence have been shown to effectively control a number of SPDE systems for reaching and stabiliziation tasks, however these policies can fail for tracking tasks. Both of these approaches are algorithmically quite similar, and may have theoretic connections if one can connect the objective in \gls{GASS} to an analogous free-energy relative entropy relationship. Aside from the differences in the resulting loss functional, another primary difference between the two approaches can be summarized by observing \cref{eq:gass_pass_grad}, wherein one passes the gradient directly to the distribution $f(\varphi;\vtheta)$ and "skips" the implicit dependence of $S(J(\rho))$ on $\theta$. This "skipped" gradient path enables one to bypass the potential discontinuities and non-differentiability of $J$, however in some sense ignores these contributions to the total gradient. These "skipped" connections may ignore important gradient information, however they offer flexibility and maintain provable convergence and convergence rate characteristics~\cite{zhou2014gradient}. 

\section*{Appendix D: QGASS Algorithm}

This section provides a brief explanation of the \gls{QGASS} algorithm. The pseudocode is presented in \cref{alg:qgass}, and represented graphically in \cref{fig:QGASS_diagram}. The inputs to the optimization include the final time ($T$), the number of iterations ($K$), the number of network parmeter rollouts ($P$), the number of trajectory rollouts ($R$), the initial state ($\rho_0$), shape function parameter ($\kappa$), initial network weights ($\varphi^{(0)}$), initial sampling distribution mean ($\vtheta^{(0)}$), and sampling distribution variance. This algorithmic pseudocode is written with multiple layers of for loops, however in implementation, these for loops were replaced by a vectorized, or batch computation that leverages parallelization of the computation. Specifically, we performed vectorized time evolution of the controlled dynamics over the trajectory rollouts, and performed CPU parallelization over prarameter rollouts.

\begin{algorithm}[H]
 \caption{Quantum Gradient-based Adaptive Stochastic Search Optimization}
 \begin{algorithmic}[1]
 \State \textbf{Function:} \textit{$\vtheta^* =$ \textbf{OptimizePolicyVars}($T$,$K$,$R$,$P$,$\rho_0$,$\kappa$,$\varphi^{(0)}$, $\vtheta^{(0)}$,$\sigma$)}
 \For {$k=0 \; \text{to} \; K$}
  \State $\mu \gets \vtheta^{(k)}$
 \For {$p=0 \; \text{to} \; P$}
 \State $\varphi_{p} \gets SampleWeights(\mu, \sigma)$
 \For {$r=0 \; \text{to}\; R$}
 \For{$t=0 \;\text{to}\; T$}
     \State $\rd W_{t,r} \gets SampleNoise()$
     \State $u_{t,r,p} \gets Policy(\rho_{t,r,p} ; \varphi_p)$
     \State $\rho_{t+1,r,p} \gets Propagate(\rho_{t,r,p}, u_{t,r,p}, \rd W_{t,r})$ via \cref{eq:dyn_controlled}
     \State $J_{t,r,p} \gets RunningCost(\rho_{t,r,p}, u_{t,r,p})$
 \EndFor
    \State $J_{r,p} \gets \sum_t J_{t,r,p} + TerminalCost(\rho_{T,r,p})$
 \EndFor
    \State $S_p \gets ShapeFunction(J_{r,p}; \kappa)$
 \EndFor
 \State $\vtheta^{(k+1)} \gets \gamma GradientStep(\vtheta^{(k)}, S_p)$ via \cref{eq:pol_ss_param_update_expectation}
 \EndFor
 \end{algorithmic}
 \label{alg:qgass}
\end{algorithm}

\section*{Appendix E: Details of the Simulation}

The \gls{QGASS} algorithm was implemented on the two qubit experiment described in the main paper. The simulation environment was created in Python and utilizes some basic functionality of the QuTip~\cite{johansson2012qutip} library, with policy networks coded using PyTorch~\cite{paszke2019pytorch}. The algorithm computation speed is numerically improved by using vectorized (or batch) computations of the simulated trajectories, and CPU parellelization for policy parameter rollouts, resulting in $\sim$20 seconds per iteration for 1000 timesteps of an Euler-Maruyama discretization of \cref{eq:dyn_controlled} with 50 rollouts and 200 policy parameter rollouts. The algorithm was run on a desktop computer with a Intel Xeon 12-core CPU with a NVIDIA GeForce GTX 1060 GPU, and used less than 10 GB of RAM.

The two qubit experiment involves a task wherein a random initial state must reach and stabilize to the symmetric maximally entangled Bell state:
\begin{equation}
   | \Psi^+ \rangle = \frac{1}{\sqrt{2}} \Big( | \downarrow_1 \uparrow_2 \rangle + |\uparrow_1 \downarrow_2 \rangle \Big),
\end{equation}
which can be written in density matrix form as
\begin{equation}
    \rho_{\text{desired}} = \left[ \begin{array}{cccc}
        0 &  0 & 0 & 0\\
        0 & 0.5 & 0.5 & 0 \\
        0 & 0.5 & 0.5 & 0 \\ 
        0 & 0 & 0 & 0
    \end{array}\right].
\end{equation}

The initial state was sampled using the `rand\_ket' qutip method, which is then transformed into a normalized density matrix. This task used a single layer fully connected linear policy network which was initialized with the method in~\cite{lecun2012efficient}. Performance of the state and control trajectories was measured by a running cost metric, given by
\begin{equation}\label{eq:qgass_cost}
    J(\rho, \vu) := \int_0^T \Big(Q_s q\big(1- \tr[\rho_{\text{desired}}\rho_\tau]\big) + \vu_\tau^\top Q_u \vu_\tau\Big) \rd \tau,
\end{equation}
where $Q_s$ is a state cost weighting and $Q_u$ is a control cost weighting. Note that this state cost metric utilizes a computationally efficient trace metric~\cite{mirrahimi2007stabilizing} as compared to the standard trace distance metric~\cite{johansson2012qutip}
\begin{align}
    \text{Tracedist}(\rho_{\text{desired}},\rho_t) &= \frac{1}{2} \text{Re} \bigg( \sum_{i=1}^{n} \sqrt{\big|\lambda_i \big[ (\rho_{\text{desired}},-\rho_t) (\rho_{\text{desired}},-\rho_t)^\dagger\big] \big|} \bigg),
\end{align}
which is substantially slower in implementation as it requires an eigenvalue decomposition at each time step. The function $q: [0,1] \rightarrow [0,\alpha]$ is an angle resolution function which is added to help resolve numerically ``close'' angular values. Recall that 
for a single qubit, the trace inner product can be thought of as measuring perpendicularity of Bloch phases. Since the cosine function is relatively flat (derivative near zero) near $0$, one may encounter bad numerical resolution near the desired minimum $1- \tr[\rho_{\text{desired}}\rho_\tau]\big) = 0$ in an n-qubit setting. The resolving function applies a logarithm transformation to improve numerical resolution, and is given by
\begin{equation}
    q(x) = \alpha \frac{\log(1 + \beta x)}{\log(1+\beta)},
\end{equation}
where $\alpha$ is the maximum of the range, and $\beta$ controls the slope by effectively changing the base of the natural logarithm.

As stated in the main paper, the cost function is passed through a differentiable and non-increasing shape function $S(\cdot): \Rb \rightarrow \Rb$. Many different shape functions can be used, as explained in greater detail in \cite{wang2021adaptive}. In our simulated experiments, we used the function
\begin{equation}
    S(J) := \exp\big(-\kappa J),
\end{equation}
where $\kappa=1.0$ affects the slope, and may be tuned for greater performance.

\end{document}